# Schrödinger's Cat – Amended*

Bernd A. Berg
berg@hep.fsu.edu

Department of Physics, The Florida State University, Tallahassee, FL 32306, USA
SCRI, The Florida State University, Tallahassee, FL 32306, USA

**Abstract**

Arguments have been raised that the system–observer cut of quantum mechanics can be shifted arbitrarily close to, or even into, the conscious observer. Here I show that this view leads to *observable* contradictions (despite our inability to control the phases of macroscopic states). For this purpose I modify and extend Schrödinger's well–known superposition of a cat in its dead and alive state. Implications for other interpretations of quantum mechanics are also discussed. My conclusion is that quantum mechanics is incomplete. The question "When does the state vector collapse?" seems to be unavoidable, has observable consequences, and is not answered by quantum mechanics.

---

\* This research was partially funded by the Department of Energy under contract DE-FG05-87ER40319.

## 1. Introduction

It is remarkable that an experimental construction[1], which the inventor himself termed "ridiculous", has over two generations caught the imagination not only of physicists. The next paragraph follows closely, though not literally, Schrödinger's own presentation in the translation of ref. [2].

*A cat is penned up in a steel chamber, along with the following diabolical device: in a Geiger counter there is a tiny bit of radioactive substance, that perhaps in the course of one hour one of the atoms decays, but also, with equal probability, perhaps none; if it happens, the counter tube discharges and through a relay releases a hammer which shatters a small flask of hydrocyanic acid. If one has left the entire system to itself for one hour, one would say that the cat still lives if meanwhile no atom has decayed. The state vector $|\Psi\rangle$ of the entire system would express this by having in it the living and the dead cat mixed or smeared out in equal parts.*

## 2. Schrödinger's Hammer

For our purposes it is necessary to explore the arrangements in details. Let us first consider the isolated state vector $|\psi(t)\rangle$ of a single radioactive nucleus. Quantum mechanics describes its time evolution through a linear operator $\mathbf{U}(t)$:

$$|\psi\rangle = |\psi(t)\rangle = \mathbf{U}(t)|\psi(0)\rangle . \qquad (1)$$

I will first work through the argument using pure states. This is not only conceptually simpler, but prevents us also from confusing the *fundamental* statistical uncertainty of quantum mechanics with the harmless statistical uncertainty of ensemble theory. The first gives rise to interference effect the latter does not. On the other hand, the system can of course never be perfectly isolated from its environment. Mixed states are addressed towards the end of the paper in a discussion of possible loopholes. The results stay unchanged. Let us assume that a *measurement* is performed at time $t = 0$, such that we know that the nucleus has not yet decayed. Afterwards it is left unobserved. The state vector evolves into

$$|\psi\rangle = |\psi_1\rangle + |\psi_2\rangle , \qquad (2)$$

where $|\psi_1\rangle$ describes the nucleus and $|\psi_2\rangle$ the decay products.

It is notable that $|\psi_2\rangle$ is now a multiparticle state, whereas $|\psi_1\rangle$ is a boundstate. The decay products are normally spread out over a space region much larger than the region where the boundstate is localized. This implies that for $t > 0$ and sufficiently large the overlap between $|\psi_1\rangle$ and $|\psi_2\rangle$ will be extremely small. It is instructive to estimate the orders of magnitude. A typical nucleus size is of the order of 10 fermi, whereas the decay products can easily be spread over a cubic meter. If this is the case,

$$|\langle\psi_1|\psi_2\rangle| = \left|\int dx \, \langle\psi_1|x\rangle\langle x|\psi_2\rangle\right| \approx \left(\frac{10^{-12}\,\text{cm}}{10^2\,\text{cm}}\right)^3 = 10^{-52} . \qquad (3)$$

Here $|x\rangle\langle x|$ defines a system of coordinate states which is complete with respect to $|\psi\rangle$. Technical details could become quite elaborate due to the involvement of multiparticle states. But it is difficult to imagine that this could change more then a few orders of magnitude in (3). For most of the discussion I am going now to neglect the overlap (3). I shall use the notation $n_i = \langle\psi_i|\psi_i\rangle$, $(i = 1, 2)$. Then $n_1(0) = 1$ and, for $t > 0$, $n_1(t)$ will monotonically decrease towards zero such that

$$n_1(t) + n_2(t) = 1. \qquad (4)$$

$n_2(t)$ is the probability that, upon observation, the nucleus did decay in the time interval $[0,t]$. In simple cases $n_1(t) = \tau^{-1} e^{-t/\tau}$, where $\tau$ is the mean-life time. This completes the discussion of a single nucleus.

In Schrödinger's construction a sample of $N \gg 1$ nuclei is used. We denote its state vector $|\Psi_N\rangle = |\Psi_N(t)\rangle$. As before, we prepare the system at time $t = 0$. For radioactive decay it is reasonable to assume (and in agreement with experiments) that the interaction between the nuclei can be neglected and one obtains

$$\mathbf{U}(t)|\Psi_N(0)\rangle = |\Psi_N\rangle = |\Psi_{N,1}\rangle + |\Psi_{N,2}\rangle \tag{5}$$

with

$$|\Psi_N\rangle = \prod_{i=1}^{N} |\psi_i\rangle, \quad |\Psi_{N,1}\rangle = \prod_{i=1}^{N} |\psi_{i,1}\rangle \text{ and } |\Psi_{N,2}\rangle = \sum_{i=1}^{N} |\psi_{i,2}\rangle \prod_{j \neq i} |\psi_{j,1}\rangle + \ldots. \tag{6}$$

Let $n_{N,i} = \langle \Psi_{N,i}|\Psi_{N,i}\rangle$, $(i = 1, 2)$. Although the equal signs in equations (6) are not exact, it is perfectly reasonable to assume that they induce an exact decomposition of the form (5), such that $|\langle \Psi_{N,1}|\Psi_{N,2}\rangle|$ is similarly small as $|\langle \psi_1|\psi_2\rangle|$ and $n_{N,2}(t)$ gives the probability that one or more nuclei decay in the time interval $[0,t]$. With $t_1 = 1$ hour $n_1(t_1) = e^{-N^{-1}\ln 2}$ will yield the desired result: the probability is 50% that one or more nuclei decayed and 50% that none decayed.

Next, as part of the described arrangement, the radioactive sample is placed inside a steel chamber. On the final level of complexity the state vector $|\Psi(t)\rangle$ of the *entire* inside system is investigated. At time $t = 0$ the system is prepared and

$$|\Psi(0)\rangle = |\Psi_N(0)\rangle \, |R(0)\rangle \,, \tag{7}$$

is considered to be a valid approximation of $|\Psi(0)\rangle$. Here $|R(0)\rangle$ is the initial state vector describing Geiger counter, relay, hammer, cat, and so on. Actually, at this point I would like to give in to the demands of animal right groups and arrange the set-up without cat and hydrocyanic acid. It will be sufficient to focus on the center of mass coordinate $\vec{x}_{cm}$ of of the hammer.

The initial factorization (7) cannot be exact, but the corrections are irrelevant. Equation (7) gives one representative out of a large class of state vectors which all lead to the same final results for the hammer's center of mass coordinate. This is quite similar to the way by which equations (6) lead to the correct result for radioactive decay. Due to the linearity of $\mathbf{U}(t)$ the radioactive decay induces

$$|\Psi(t)\rangle = \mathbf{U}(t)|\Psi(0)\rangle = |\Psi_1(t)\rangle + |\Psi_2(t)\rangle, \tag{8}$$

corresponding to the first term versus the second in equation (5). Let us be precise about the properties of $|\Psi_1\rangle$ and $|\Psi_2\rangle$. We assume that the overlap $|\langle \Psi_1|\Psi_2\rangle|$ can be neglected, at least for $t > 0$ sufficiently large, and then

$$\langle \Psi_i(t)|\vec{x}_{cm}|\Psi_i(t)\rangle = \vec{x}_{cm,i}, \ (i = 1, 2)\,. \tag{9}$$

Here $\vec{x}_{cm,1}$ is the initial center of mass position of the hammer and $\vec{x}_{cm,2}$ is a mean over hammer positions (including movements) which would have killed the cat. In particular for $t = t_1 = 1$ hour $0.5 = N_1(t_1) = N_2(t_1)$ with $N_i = \langle \Psi_i|\Psi_i\rangle$, $(i = 1, 2)$.

With $|R(t)\rangle = \mathbf{U}(t)|R(0)\rangle$, $(t > 0)$ it should be noted

$$|\Psi_1(t)\rangle = |\Psi_{N,1}(t)\rangle \, |R(t)\rangle \tag{10}$$

can be considered as a valid approximation of $|\Psi_1\rangle$, whereas

$$|\Psi_2(t)\rangle \ne |\Psi_{N,2}(t)\rangle |R(t)\rangle, \tag{11}$$

as $\langle\Psi_{N,2}(t)|\langle R(t)|\vec{x}_{cm}|R(t)\rangle|\Psi_{N,2}(t)\rangle = \vec{x}_{cm,1}$ and not $\vec{x}_{cm,2}$. Hence, the the unitary time evolution (8) involves non-trivial interactions of a macroscopic number of degrees of freedom and can certainly not be calculated explicitly. How can we then claim $\langle\Psi_2|\vec{x}|\Psi_2\rangle = \vec{x}_{cm,2}$ and $\langle\Psi_1|\Psi_2\rangle \approx 0$? The reason is a consistency argument, which will be a bit lengthy.

Initially $|R(0)\rangle$ is prepared as a *measurement device*. Let us for the moment assume that someone observes the Geiger counter directly, for instance by sitting next to it and keeping records. Assuming 100% efficiency, the observer will report that the Geiger counter fires with certainty, whenever a charged particle of sufficiently high momentum passes through it. Subsequently, the apparatus acts deterministically and moves the hammer into a new position. For this part the classical limit of quantum mechanics is a valid approximation. The role of the steel chamber is to shield against cosmic and possibly other outside radiation. In this way the observer can make sure that the inside radioactive sample has been the cause of those actions. Experimentally, these statements can be tested in various ways. (i) Instead of the radioactive sample one may place a sample of stable nuclei in the Geiger counter. The counter tube will then never discharge or, otherwise, allows to estimate the correction to perfect shielding. (ii) One may include a small accelerator inside the steel chamber. Then experiments may be performed where one makes sure that, at time $t_0$, just one appropriate particle enters the Geiger counter. The counter tube should then always discharge or, otherwise, allows to estimate the correction to the assumed 100% efficiency.

Consider the state vector $|f(t_0,\vec{x}_i)\rangle$ of a single particle emitted at time $t_0$ at position $\vec{x}_i$ inside the Geiger counter. Ignoring spin complications, $|f(t_0,\vec{x}_i)\rangle$ is something like

$$|f(t_0,\vec{x}_i)\rangle = |f(t,\vec{x};t_0,\vec{x}_i)\rangle = N^{-1} \int dt' \, d^3x' \, e^{-\frac{(t'-t_0)^2}{(\Delta t')^2}} e^{-\frac{(\vec{x}'-\vec{x}_i)^2}{(\Delta x')^2}} \int d\Omega \int_0^\infty (k_0)^2 \, dk_0 \, e^{-\frac{(k_0-\overline{k})^2}{(\Delta k)^2}}$$

$$\times \left[\theta(t-t')\,\theta((\vec{x}-\vec{x}')^2 - a^2) + \theta(t'-t)\,\theta(a^2 - (\vec{x}-\vec{x}')^2)\, e^{-\frac{(t-t')^2}{(\Delta t)^2}} e^{-\frac{(\vec{x}-\vec{x}')^2 - a^2}{(\Delta a)^2}}\right]$$

$$\times \int d^3k \, e^{i(t-t')\sqrt{m^2+(\vec{k}-\vec{k}_0)^2}} e^{-i(\vec{x}-\vec{x}')\cdot(\vec{k}-\vec{k}_0)} c^\dagger_{\vec{k}}|0\rangle. \tag{12}$$

Here $\overline{k}$ is the absolute value of a typical momentum of the particle and $\vec{k}_0 = \vec{k}_0(k_0,\Omega)$, $k_0 = |\vec{k}_0|$. The various widths should be self-explaining. With the normalization factor $N^{-1}$ chosen such that $\langle f(t_0,\vec{x}_i)|f(t_0,\vec{x}_i)\rangle = 1$ experimental facts tell us that the Geiger counter will discharge with probability one at a time close to $t_0$. The same facts teach us that the detailed structure of the state vector $|f(t_0,\vec{x}_i)\rangle$ is irrelevant, but it is instructive to have one in mind. For a single, unobserved nucleus at position $\vec{x}_i$ the decay products may be represented as

$$|\psi_{i,2}(t)\rangle = \sqrt{n_2(t)} \int_0^t dt_0 \, |f(t_0,\vec{x}_i)\rangle \, |g(t_0,\vec{x}_i)\rangle, \tag{13}$$

where $|g(t_0,\vec{x}_i)\rangle$ represents the other decay products, correlated with $|f(t_0,\vec{x}_i)\rangle$ through energy–momentum and other conservation laws. Assume now continuous observation of the single nucleus through Geiger counter plus observer. With no decay observed, $|\psi_{i,2}\rangle$ becomes

$$|\psi_{i,2}(t)\rangle = \sqrt{n_2(\tau)} \int_{t-\tau}^t dt_0 \, |f(t-\tau,\vec{x}_i)\rangle \, |g(t-\tau,\vec{x}_i)\rangle, \tag{14a}$$

where $\tau$ is a delay time, depending on details of the Geiger counter. The state vector $|\psi\rangle$ is still given by (2), but the normalization of $|\psi_1\rangle$ is continuously redefined through $n_1(t) = 1 - n_2(\tau)$. Now under observation of a decay at time $t = t_0$ (no position measurement) the following transition happens

$$|\psi_i\rangle = |\psi_{i,1}\rangle + |\psi_{i,2}\rangle \;\rightarrow\; |f(t_0 - \tau, \vec{x}_i)\rangle |g(t_0 - \tau, \vec{x}_i)\rangle \,. \tag{14b}$$

The branch $|\psi_{i,1}\rangle$ disappear entirely and $|\psi_{i,2}\rangle$ gets reduced to a particular case. It is straightforward to extend the discussion to measurement of $|\Psi_N\rangle$ defined by equation (6). Equations (14) define the *reduction* or *collapse* of the state vector.

The time evolution with $\mathbf{U}(t)$, see equations (1), (5) and (6), versus equations (14) are the two different dynamical laws of quantum mechanics which were first clearly formulated by von Neumann[3]. The linear, smooth and unitary evolution with $\mathbf{U}(t)$ is claimed to be valid for isolated or closed systems. In contrast, equations (14), in particular (14b), describe an abrupt probabilistic transition into a new state vector and are claimed to be valid for measurements. The *system–observer cut* is located wherever the transition between the two laws takes place and will be discussed in the next section.

### 3. The Quantum Measurement Problem and Interpretations I

The quantum measurement problem is that the time evolution with $\mathbf{U}(t)$ and equations (14) are mathematically inconsistent with one another. However, whether the time evolution $\mathbf{U}(t)$ will give physically *observable* disagreement with the procedure (14) is less clear. There is considerable interpretational freedom about what constitutes a measurement and what an isolated or closed system. The question is, can the cut be made without getting into contradictions with the unitary time evolution $\mathbf{U}(t)$? Quantum mechanics has produced a wealth of interference phenomena. For instance double slit interference has recently been observed using atoms[4]. To avoid disagreement with such phenomena, the system–observer cut has to stay sufficiently far away from the microscopic world. However, concerning the other limit, it seems to be a widespread opinion that the cut can be moved arbitrarily close to the conscious observer without encountering any observable discrepancy. Let me explain this.

In our example, a reasonable conjecture is certainly that the cut takes place in the Geiger counter (or before), because this is in agreement with direct observation of the Geiger counter. Is there now an observable discrepancy with assuming continued unitary time evolution of the entire inside system? To be definite, let us remove our observer from the steel chamber and arrange a closed system. It is certainly no problem to include power supply etc. and to provide shielding against curious outside observers, such that there is *in practice* no other way to find out about the position of the hammer, then by opening the steel chamber. No time records about the discharge of the counter tube, movement of the hammer, etc. are taken. It is questionable whether this shielding can ever be made so perfect that also theoretically no outside observation is possible, but this is a minor point as we will see in the discussion of loopholes.

As before, we are not confined to just one experiment. The observer may open the steel chamber at variant times, place a non-radioactive sample in it, or perform experiments by preparing the little accelerator (still inside). By placing a non–radioactive sample $|\Psi_{N,1}\rangle$ inside, we find experimental consistency with the obvious time evolution (10). By using our accelerator to generate $|\Psi_2(0)\rangle = |f(0, \vec{x}_i)\rangle |R(0)\rangle$, we find a non-trivial time development for $|\Psi_2\rangle$, which for $t > 0$ (large enough) always leads to the observation

$\langle\Psi_2|\vec{x}_{cm}|\Psi_2\rangle = \vec{x}_{cm,2}$, as demanded by equation (9). Our radioactive sample is now just a linear superposition of $|\Psi_{N,1}\rangle$ and $|\Psi_{N,2}\rangle$, where $|\Psi_{N,2}\rangle$, is obtained by inserting $|\psi_{i,2}\rangle$ given by (13) in (6). As the operator $\mathbf{U}(t)$ is linear, we inevitably arrive at the result (8). This is Schrödinger's observation that the linear operator $\mathbf{U}(t)$ will not collapse the wave function into either one of the possibilities $|\Psi_1\rangle$ or $|\Psi_2\rangle$, but instead ends up with a superposition of both. Only by the act of *measurement* (whatever this is) the wave function will collapse. The result is, of course, in disagreement with the state vector which the inside observer would report.

Apparently, over two generations, most physicists were not too seriously worried by this. A prevailing, pragmatic point of view emphasizes that quantum mechanics just works: Opening the steel chamber after one hour, the observer will find the hammer with 50% probability at position $\vec{x}_{cm,1}$, with almost 50% probability at $\vec{x}_{cm,2}$ and with a small likelihood it is caught moving. This is just what the state vector (8) of the inside system predicts. Beyond the prevailing view different interpretations are debated. The convinced pragmatist perceives this discussion as *metaphysics*, where metaphysics is defined as an issue of interpretation only without observable consequences. Of course, this did not stop the debate. Nine interpretations of Schrödinger's cat problem were recently reviewed[5]. At this point I like to emphasize two of them.

The disagreement of the state vectors under different observation procedures may be regarded as a failure of *ontology* (theory of being). The interpretation of Bohr[6,7], and in a limited sense the Copenhagen interpretation, reduced physics to *epistemology* (theory of knowledge). It is rejected that the state vector represents physical reality. What matters is to predict observations correctly. In philosophy this is reflected by the influential positivistic school. This leads to Einstein's question "Is the moon there when nobody looks?", which was restated[8] after the experimental[9] failure of local hidden variable theories[10].

Some authors[11,12] have speculated that the conscious observer may be responsible for the ultimate collapse of the wave function. This allows a dynamical[13] chain between observers and observed. It assumes that, empirically, we know only that a collapse has taken place when the information arrives at the final link. If there are several observers involved in a chain, known as "Wigner's friends", each one describes consistent results.

### 4. An Interference Experiment

My modification of Schrödinger's arrangement is intended to show that the collapse of the state vector has to happen sometime before the hammer moves, and that this is implied by physics, not metaphysics.

The arrangement is depicted in the figure. $G$ denotes the Geiger counter with radioactive material placed in it as before. There are now two changes. First, hammer etc. are replaced by a system consisting of a laser, two mirrors and a screen. Second, after one hour, the observer is only allowed some partial insight. Information is transferred to him about the intensity pattern which the laser produces on the indicated screen. He will have no knowledge about anything else going on inside. Laser and mirrors are assembled as follows. Mirror $M_1$ is fully silvered and in a fixed position. Mirror $M_2$ can switch from an up to a down position and vice versa. When $M_2$ is in its up position, the laser beam illuminates directly the screen. When a fully silvered mirror $M_2$ is placed in the down position, the beam is directed towards $M_1$ and from there to the screen. Finally, when a half silvered mirror $M_2$ is placed in the down position, as indicated in the figure, a characteristic interference pattern will result on the screen.

We start the experiment at time $t = 0$. Initially the system is equipped with a fully silvered mirror $M_2$ in down position. Through the relay, which previously triggered the hammer, the mirror $M_2$ will be moved into up position when the counter tube discharges. After one hour, which intensity pattern will be transferred to the outside observer? As before we denote by $|\Psi(t)\rangle$ the state vector of the entire inside system and it will develop into the superposition (8). Let $\vec{x}_{cm}$ now denote the center of mass coordinate of mirror $M_2$. Equation (9), indicating the two final positions, holds again. It is even more instructive to project out the corresponding wave function. As in our discussion of the particles triggering the Geiger counter, details are not expected to matter. A reasonable model wave function is then ($i = 1, 2$ and $\mathbf{P}$ projection operator)

$$\psi_i(t, \vec{x}) = \langle (t, \vec{x}) \mathbf{P} | \Psi_i \rangle = N^{-1} \sqrt{N_i(t)} \int_0^t dt_1 \int_0^{t_1} dt_0 \frac{\exp\left[ -i\phi_i - \frac{(\vec{x} - \vec{y}_i(t_1 - t_0))^2}{(2\Delta x)^2} + \frac{it(\vec{x} - \vec{y}_i(t_1 - t_0))^2}{8m(2\Delta k)^{-4} + 2t^2/m} \right]}{[(2\Delta k)^{-2} + it/(2m)]^{3/2}}. \quad (15)$$

Here $\Delta x = \sqrt{(2\Delta k)^{-2} + t^2(2\Delta k)^2/(2m)^2}$, $\vec{y}_1(\tau) = \vec{x}_{cm,1}$ and $\vec{y}_2(\tau)$ is a smooth, monotone function such that $\vec{y}_2(0) = \vec{x}_{cm,1}$ and $\vec{y}_2(\tau) = \vec{x}_{cm,2}$ for $\tau \geq \tau_1$, with $\tau_1 \ll 1$ hour. The normalization factor $N^{-1}$ is chosen such that $\int d^3x \, \overline{\psi}_i(\vec{x}) \psi_i(\vec{x}) = N_i(t)$ holds. It is easily seen that with increasing $t$ the overlap $\int d^3x \, |\overline{\psi}_1(\vec{x}) \psi_2(\vec{x})|$ rapidly approaches zero. (It should be noted that this sets in for very small separations of $\vec{y}_1$ versus $\vec{y}_2$, due to the unknown phases $\phi_i$ and the rapid oscillatory behavior of the $it(\vec{x} - \vec{y}_i)^2/(8m(\Delta k)^{-4} + 2t^2/m)$ term.) Assume that the mirror has a mass of 1 gram, then the wave function stays perfectly localized, when it was initially. For instance, it would take almost three thousand years to double an initial spread of the order of one Bohr radius $a_0$. The corresponding momentum uncertainty $\Delta k = \hbar/(2a_0)$ would be smaller then $10^{-16}$ [g cm/s]. This is just to reiterate my earlier statement that, once the counter tube discharged, the classical limit of quantum mechanics is valid. The dominant fluctuations will be thermal. For each branch the mirror wave function behaves deterministically. Now, let $\vec{x}$ be a point on the screen. Consider photons $\gamma_{\text{in}}$, originating from the laser. Their contribution to the light intensity at $\vec{x}$ is

$$I(\vec{x}) = |\langle \gamma(\vec{x}) \mathbf{P} | \Psi_1 \rangle \langle \Psi_1 | \mathbf{P} \gamma_{\text{in}} \rangle + \langle \gamma(\vec{x}) \mathbf{P} | \Psi_2 \rangle \langle \Psi_2 | \mathbf{P} \gamma_{in} \rangle |^2. \quad (16)$$

As each of the intermediate states $|\Psi_1\rangle\langle\Psi_i|$, ($i = 1, 2$) behaves classically, we know from the arrangement of classical mirrors that they will cause relative phase shift in the arriving light. Ignoring corrections which follow from the detailed form (15) of $\psi_2(t, \vec{x})$, the effect after $t_1 = 1$ hour will be the same as that of a half-silvered mirror in down position. The corresponding characteristic interference pattern should be transferred.

I think, one does not need to perform the experiment, to know that this will not be the outcome. Mirror $M_2$ will either be in the up or in the down position (ignoring the short time period in between) and those intensity patterns will be transferred, each with 50% probability. The state vector collapsed before mirror $M_2$ got moving. The moon, or at least mirror $M_2$, is there when nobody looks.

Within conventional quantum mechanics the explanation has to be that screen, mirrors, laser, relay and Geiger counter altogether constitute a measurement device. Therefore, the intensity pattern on the screen allows us to conclude whether a radioactive decay has happened or not. Further, and most important, we are not allowed to confuse a measurement device with a quantum object. Applying the linear operator $\mathbf{U}(t)$ to the joint system gives manifestly wrong results. A measurement device collapses wave functions, that's what it is made for. The bizarre wave function $\psi = \psi_1 + \psi_2$ with $\psi_i$ given by (15) is a consequence of applying unitary time evolution far beyond its range of validity. A macroscopic body, like mirror $M_2$, is there when nobody looks (*i.e.* observes or measures its position directly). The mirror $M_2$ refuses to be in a superposition of up and down, and to produce the corresponding interference pattern on the screen.

The question is then, what constitutes a measurement device and what a quantum system? Quantum mechanics is incomplete as it contains neither unambiguous rules which allow to distinguish a measurement device from a quantum system, nor a generally accepted theory of measurement.

Or, are there loopholes in the argument? Can we rescue the assumption that unitary time evolution $\mathbf{U}(t)$ can be consistently applied to macroscopic systems, including measurement devices? Let me address two issues shortly: (i) Imperfect shielding against outside observations. (ii) Mixed states due to interactions with the environment (inside and outside).

(i) We have seen that the state vector of a *practically* unobservable macroscopic system, itself a measurement device, has to collapse at some point. A theory of measurement could claim that the state vector collapses whenever it is in principle observable. It remains then to define what in principle observable means, and the theory of state vector collapse is complete. Nothing is gained by just hiding the problem behind undefined words.

(ii) An often voiced opinion is that interactions with macroscopic bodies (in particular the never exactly controllable environment) cause decoherence such that no measurable interference effects are left over. It is correct, and made explicit by the wave function (15), that mirror $M_2$, or originally the cat, cannot interfere with itself. However, this does not relieve us from *independently* postulating the disappearance of redundant branches. Otherwise, we end up with macroscopic contradictions. In fact, in my arguments I made extensively use of the early decoherence of the quantum system, see equation (3). This allowed to keep the branches of (8) convincingly apart and to use the classical limit to create an interference effect. Our state vector was typical for all conceivable states. This makes the analysis of mixed states an easy exercise. Instead of the pure state (8) we consider the density matrix $\rho = \sum p^n |\Psi^n\rangle\langle\Psi^n|$. Each state decomposes like (8): $|\Psi^n\rangle = |\Psi_1^n\rangle + |\Psi_2^n\rangle$, with practically no overlap between $|\Psi_1^n\rangle$ and $|\Psi_2^n\rangle$. Equation (9) becomes $\vec{x}_{cm,i} = Tr(\rho_i \vec{x}_{cm})$, with $\rho = \sum p^n |\Psi_i^n\rangle\langle\Psi_i^n|$, $(i = 1, 2)$, and $I(\vec{x}) = |\gamma_1(\vec{x}) + \gamma_2(\vec{x})|^2$ with $\gamma_i(\vec{x}) = Tr(|\gamma_i n\rangle\langle\gamma(\vec{x})| \rho_i)$. We end up with two branches of mixed states with the same properties as before. One has to postulate the disappearance of one of them to avoid macroscopic contradictions. (Note that we are interested in the behavior of a *single* system.) Just the opposite of decoherence would be needed. If environmental interactions could mix up the branches again, and destabilize one in favour of the other, the interference effect would be in trouble. Due to the linear time evolution $\mathbf{U}(t)$ this looks inconceivable.

### 5. Interpretations II

To be consistent with the outlined experiment, quantum mechanics interpretations have to fulfil two requirements:

(i) The state vector (2) or (5) or (8) collapses, through whatever mechanism, into one of the two described branches.

(ii) This collapse has to happen sufficiently fast, such that the two branches cannot be put together to

produce interference effects through superposition of macroscopic bodies.

Let me discuss a few interpretations in this light.

In the many worlds scenario[14,15] one branch of our state vector would just disappear into another world. This provides us with the instructive picture that the considered interference effect results from superimposing two worlds which in fact are separated. Although this may look appealing, the real lesson of our construction is that those worlds do not decouple automatically. It is an independent postulate that the different world branches cannot be brought back into interaction with one another. The scenario is then physically identical to postulating a state vector collapse at each branch point. Whether the superfluous branch just disappear or lives on in another world becomes a question of metaphysics, and economy of thinking does strongly favour a conventional collapse approach. A many worlds interpretation distracts from the fact that the state vector reduction remains as badly understood as before.

Recent interpretations of quantum mechanics tend towards ontology which, by its very nature, avoids paradoxical situations like the one outlined. Decoherence and consistent histories have emerged as a central concept, see ref.[16,17,18] for reviews. It is well known[17] that decoherence is logically independent from disappearance of the superfluous branches. My example seems to show it needs to evolve into a state vector reduction theory or at least has to specify reduction criteria. Some state vector collapse ideas[19] and models[20] are discussed in the literature, see there for further references.

## 6. Outlook

The early decoherence (3) suggests to me that, in particular situations, collapse may already happen on the few particle level. For instance, after ten minutes an unobserved neutron should be in a superposition of approximately half neutron and half decay products. It would be very interesting to know, whether it really is. The involved time and energy scale makes this questionable, at least to the author. According to the lines of thought in this paper, the idea would be to use the two branches for producing an interference effect of a third object with itself. It is beyond the scope of this article to discuss whether such experiments could be realistically designed on the microscopic level, but I have some hope that this paper may inspire experimentally more challenging tasks than the one outlined.

**Acknowledgements:** I would like to thank Wolfgang Beirl for a number of very useful comments and Tai Tsun Wu for encouragement.

## References

1) E. Schrödinger, Naturwissenschaften **23**, 807, 823, 844 (1935).
2) J.A. Wheeler and W.H. Zurek (editors), *Quantum Theory and Measurement*, Princeton University Press, Princeton 1983.
3) J. von Neumann, *Mathematische Grundlagen der Quantenmechanik*, Springer, 1932.
4) M.S. Chapman et al., Phys. Rev. Lett. **74**, 4783 (1995).
5) Z. Schreiber, preprint, quant-ph/9501014.
6) N. Bohr, Phys. Rev. **15**, 696 (1935), reprinted in ref.[2], p.145-151.

7) E. Scheibe, *The Logical Analysis of Quantum Mechanics*, Volume 56 of the *International Series of Monographs in Natural Philosophy*, Pergamon Press, 1973.

8) N.D. Mermin, Physics Today **38(4)**, 38 (1985).

9) P.R. Tapster, J.G. Rarity, and P.C.M. Owens, Phys. Rev. Lett. **73**, 1923 (1994); Y.H. Shih and C.O. Alley, Phys. Rev. Lett. **61**, 2921 (1988); W. Pierre, A.J. Duncan, H.J. Beyer, and H. Kleinschoppen, Phys. Rev. Lett. **58**, 1790 (1985); A. Aspect, J. Dalibard, and G. Roger, Phys. Rev. Lett. **49**, 1804 (1982); E.S. Fry and R.C. Thomson, Phys. Rev. Lett. **34**, 465 (1976); S.J. Freedman and J.F. Clauser, Phys. Rev. Lett. **28**, 938 (1972).

10) A. Einstein, B. Podolsky, and N. Rosen, Phys. Rev. **47**, 777 (1935); J.S. Bell, Physica **1**, 195 (1964); J.D. Franson, Phys. Rev. Lett. **62**, 2205 (1988).

11) F. London and E. Bauer, in ref.[2], p.217-259.

12) E.P. Wigner, in ref.[2], p.168-181.

13) H.D. Zeh, Found. Phys. **3**, 109 (1973).

14) H. Everett, Rev. Mod. Phys. **29**, 454 (1957), reprinted in ref.[2], p.315-323.

15) B.S. DeWitt and R.N. Graham (editors), *The Many Worlds Interpretation of Quantum Mechanics*, Princeton University Press, Princeton 1973.

16) W.H.Zurek, Physics Today **44(10)**, 36 (1991).

17) H.D. Zeh, preprint, quant-ph/9506020.

18) F. Dowker and A. Kent, preprint, gr-qc/9412067.

19) R. Penrose, *The Emperor's New Mind*, Penguin, New York 1991.

20) P. Pearle and E. Squires, preprint, quant-ph/9503019.

**Figure Caption**

Schrödinger's cat – amended. $G$ denotes a Geiger counter in which radioactive material is placed. $M_1$ and $M_2$ are fully silvered mirrors. Other details are explained in the text.

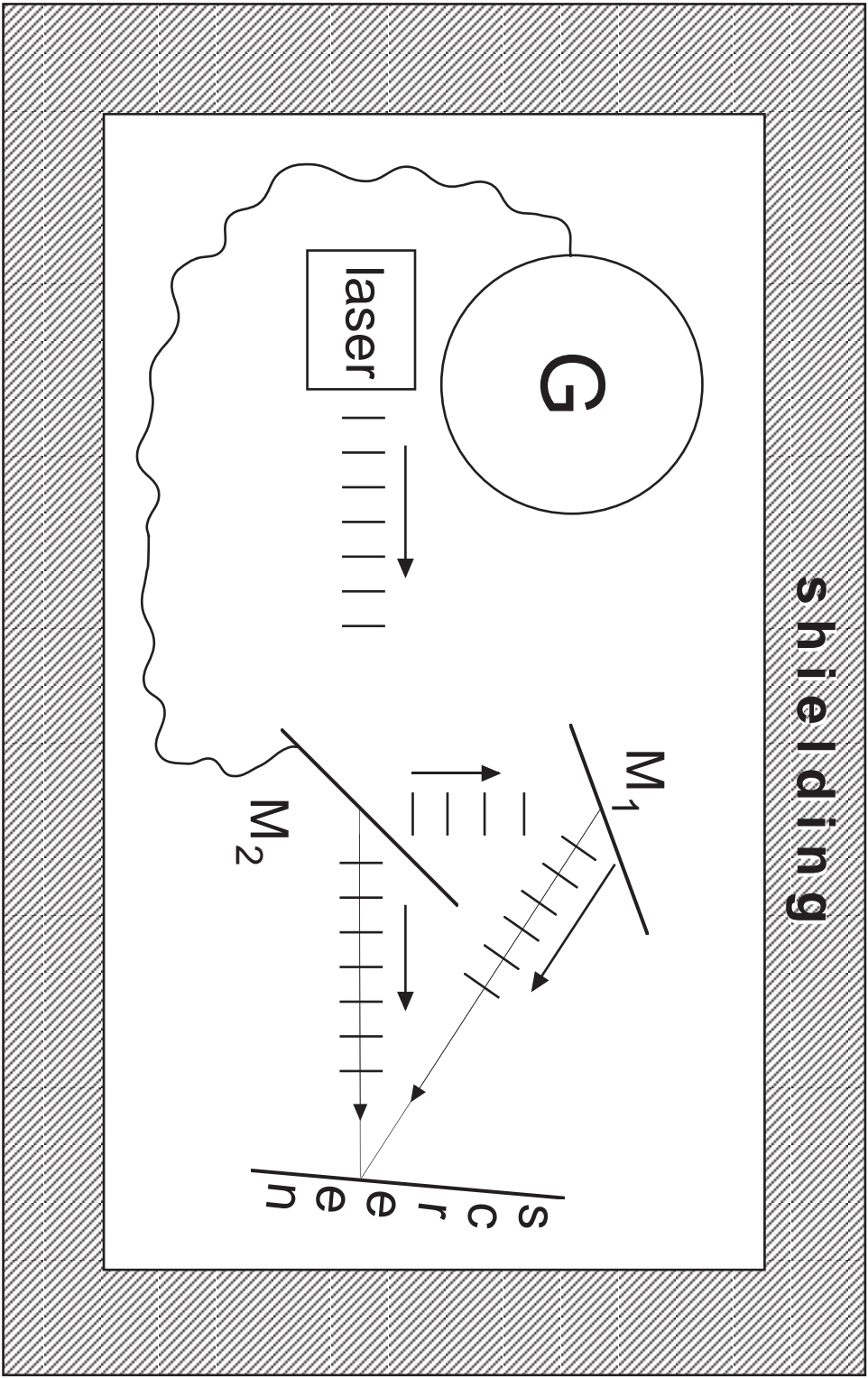